\documentclass[aps,pra,twocolumn,superscriptaddress]{revtex4-1}

\usepackage{amsmath}
\usepackage{amssymb}

\usepackage{xcolor}

\def\red{\color{red}}

\def\blue{\color{blue}}

\usepackage{bm}
\usepackage{graphicx}
\usepackage{hyperref}
\usepackage[caption=false]{subfig}
\usepackage{adjustbox}

\newcommand*\diff{\mathop{}\!\mathrm{d}}
\newcommand*\Diff[1]{\mathop{}\!\mathrm{d^#1}}

\usepackage{minitoc}

\AtBeginDocument{\let\oldcontentsline\contentsline}
\newcommand{\notoccontentsline}[4]{\oldcontentsline{}{}{}{}}
\newcommand{\droptocpage}{\addtocontents{toc}{\let\protect\contentsline\protect\notoccontentsline}}
\newcommand{\incltocpage}{\addtocontents{toc}{\let\protect\contentsline\protect\oldcontentsline}}

\newcommand{\cmmnt}[1]{\ignorespaces}

\begin{document}


\title{Measuring Electromagnetic and Gravitational Responses of Photonic Landau Levels}

\author{Nathan Schine}
\affiliation{James Franck Institute and the Department of Physics, University of Chicago, Chicago, Illinois, 60637}
\author{Michelle Chalupnik}\thanks{Present Address: Department of Physics, Harvard University, Cambridge, MA 02138}
\affiliation{James Franck Institute and the Department of Physics, University of Chicago, Chicago, Illinois, 60637}
\author{Tankut Can}
\affiliation{Initiative for the Theoretical Sciences, The Graduate Center, CUNY, New York, New York, 10012}
\author{Andrey Gromov}
\affiliation{Kadanoff Center for Theoretical Physics and Enrico Fermi Institute, University of Chicago, Chicago, Illinois, 60637}
\author{Jonathan Simon}
\affiliation{James Franck Institute and the Department of Physics, University of Chicago, Chicago, Illinois, 60637}
\date{\today}
\begin{abstract}
The topology of an object describes global properties that are insensitive to local perturbations. Classic examples include string knots and the genus (number of handles) of a surface: no manipulation of a closed string short of cutting it changes its ``knottedness''; and no deformation of a closed surface, short of puncturing it, changes how many handles it has. Topology has recently become an intense focus of condensed matter physics, where it arises in the context of the quantum Hall effect~\cite{vonKlitzing1980IQH} and topological insulators~\cite{chen2009experimental}. In each case, topology is defined through invariants of the material's bulk~\cite{TKNN1982,kane2005z,bradlyn2017topological}, but experimentally measured through chiral/helical properties of the material's edges. In this work we measure topological invariants of a quantum Hall material through local response \emph{of the bulk}: treating the material as a many-port circulator enables direct measurement of the Chern number as the spatial winding of the circulator phase; excess density accumulation near spatial curvature quantifies the curvature-analog of charge known as mean orbital spin, while the moment of inertia of this excess density reflects the chiral central charge. We observe that the topological invariants converge to their global values when probed over a few magnetic lengths $l_B$, consistent with intuition that the bulk/edge distinction exists only for samples larger than a few $l_B$. By performing these experiments in photonic Landau levels of a twisted resonator~\cite{schine2016synthetic}, we apply quantum-optics tools to topological matter. Combined with developments in Rydberg-mediated interactions between resonator photons~\cite{jia2017strongly}, this work augurs an era of precision characterization of topological matter in strongly correlated fluids of light.
\end{abstract}

\maketitle

\droptocpage

\begin{figure*}
\includegraphics[width=.93\textwidth]{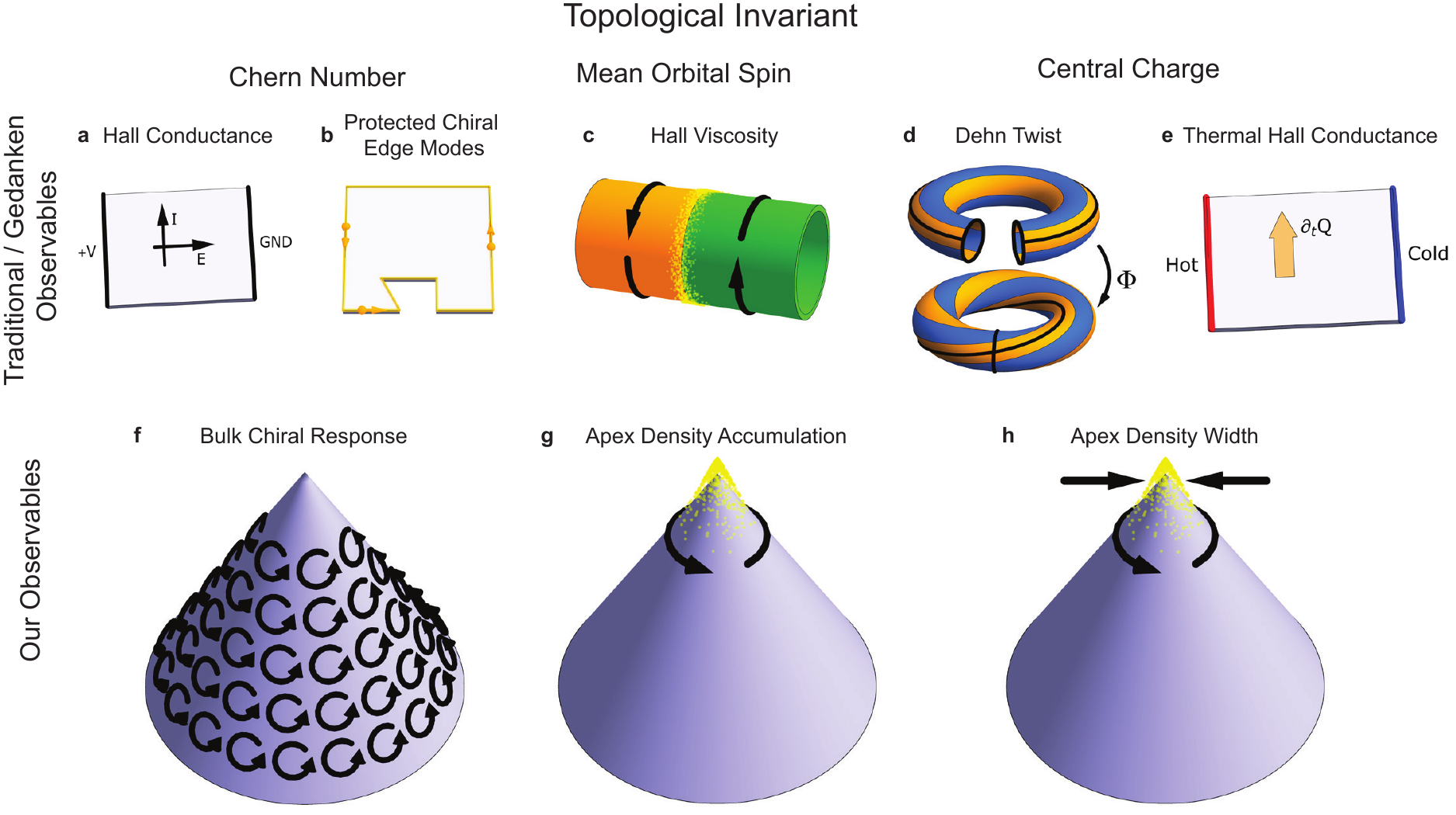}
\caption{\textbf{Topological Invariants and their Associated Observables.} \textbf{(a)} In solids, a band's Chern number is typically obtained through a Hall conductance measurement, whose precise quantization arises \textbf{(b)} from the presence of chiral disorder-protected edge-channels. \textbf{(c)} The mean orbital spin is the orbital angular momentum carried by particles in a quantum Hall fluid; it gives rise to Hall viscosity, a dissipationless transverse diffusion of momentum. It may be measured at a shear interface, where the fluid flows with two different speeds (orange and green regions); the Hall viscosity determines the accumulation/depletion of particles at this interface. \textbf{(d)} The central charge is the third topological invariant characterizing quantum Hall fluids \cmmnt{XXX but what IS it? need this hear to complete the parallel structure!}; it is most simply understood as a manybody phase-accumulation in response to a ``Dehn twist'' of the torus on which a fluid resides, and most directly measured through  \textbf{(e)} the thermal Hall conductance, where the heat flow, $\partial_t Q$, is perpendicular to a temperature gradient. \textbf{(f-h)} In our system, we measure all three topological invariants through newly accessible observables. \textbf{(f)} Working in flat space away from the cone apex, we measure a bulk chiral phase response (Fig. \ref{Figure:MeasurementScheme}), to extract the Chern number (Fig. \ref{Figure:ChernNumber}). \textbf{(g)} We measure the accumulation of particles at the cone apex and \textbf{(h)} associated orbital angular momentum (via the moment of inertia of the apex density), to extract the mean orbital spin and central charge (Fig. \ref{Figure:ldos}).\label{Figure:Observables}}
\end{figure*}

\begin{figure*}[t]
\includegraphics[width=\textwidth]{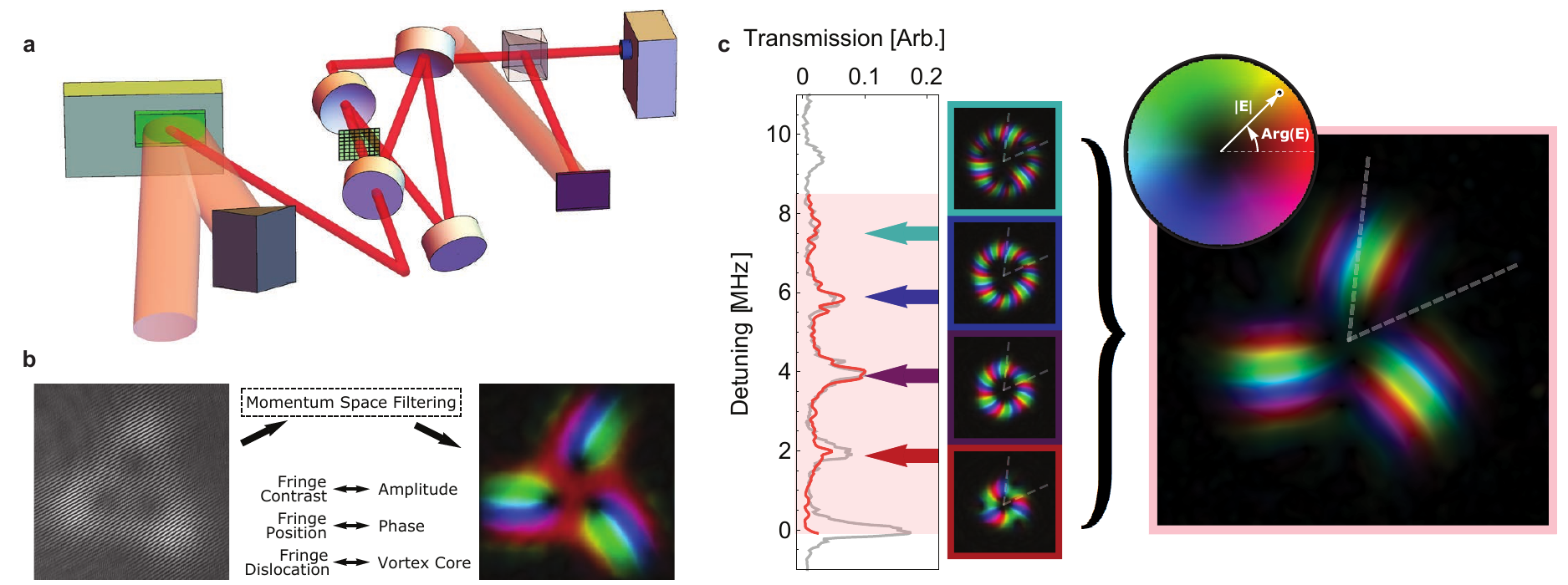}
\caption{\label{Figure:Setup}\textbf{Holographic Reconstruction of Band-Projectors.} Holographic beam shaping allows injection of arbitrary light fields into our photonic quantum Hall system, while heterodyne imaging of the cavity leakage field enables full complex-valued electric-field reconstruction of the system's response. \textbf{(a)} A 780 nm laser field is directed onto a digital micromirror device (DMD, green) and diffraction off of the DMD's hologram is directed into the non-planar resonator. Light leaking from the cavity through one of its mirrors is split on a 50:50 beamsplitter and directed to a photodiode (blue), and camera (purple) that images the transverse plane at the waist of the cavity (green grid). A few percent of the initial input light forms a reference beam that is also directed onto the camera but at a significant angle relative to the resonator output to enable heterodyne imaging akin to optical holography. \textbf{(b)} The plane wave reference beam interferes with the cavity output to produce an image (left) where the fringe contrast provides field amplitude information, and fringe position provides field phase information. This information is extracted from the images via a filtering scheme in momentum space (see SI \ref{SI:EfieldReconstruction}), providing the cavity mode electric field profile (right). \textbf{(c)} The projectors used to extract the Chern number are measured by injecting a (magnetically) translated Gaussian beam and integrating (via long camera exposure) the heterodyned cavity response while sweeping the laser frequency across the Landau level. This procedure is robust to potential disorder that broadens the Landau level so long as the disorder is not strong enough substantially admix other Landau levels. To demonstrate this robustness we apply weak harmonic confinement: the individual eigenmodes are then Laguerre-Gaussian rings (small boxes \& gray trace); a displaced Gaussian beam has significant overlap with only a few of these modes (red trace), but integrating across the relevant frequency band (pink) yields a \emph{localized} response (pink box) from which we extract the projector; this is because the holographic reconstruction effectively integrates the complex electric \emph{field} leaking from the cavity, rather than its \emph{intensity}, resulting in constructive interference of the various modes along the vertical dashed lines, and destructive interference along the diagonal lines (see SI~\ref{SI:EfieldReconstruction}). This field-integration is insensitive to potential disorder that broadens the band.}
\end{figure*}

Topological phases of matter, which cannot be characterized by the spontaneous breaking of a local symmetry, have revolutionized modern condensed matter physics and materials science~\cite{HaldaneNobelLecture,Chiu16}. Such phases are so named because they possess global invariants which are insensitive to material imperfections. These invariants have found applications from the redefinition of the unit of electrical resistance to error-resilient spintronics~\cite{moore2010birth} and quantum computation~\cite{kitaev2003fault}.

Constructed as integrals of a ``curvature'' over a closed parameter space, these invariants are each defined as a \emph{global} property resulting from the integral of a \emph{local} property, akin to the relationship between the (local) Gaussian curvature of a surface and the (global) Euler characteristic which determines the number of handles of the surface. In the integer quantum Hall effect, integration of the Berry curvature over the Brillouin zone (momentum space) defines an invariant called the first Chern number ~\cite{TKNN1982,Hatsugai1993}.  Two additional topological invariants, the mean orbital spin and central charge, are defined similarly to the Chern number, but over even more abstract parameter spaces~\cite{Avron95,can2016emergent}\footnote{An attractive alternative approach is to employ a topological quantum field theory. Although unnecessary for integer quantum Hall, this allows to argue the quantization and invariance of $\nu$, $\bar{s}$, and $c$ on simpler principles and applies equally well to integer and fractional states.}\footnote{We use central charge and chiral central charge interchangeably in this article.}.

Understanding the physical significance of the invariants characterizing topological matter remains a challenge. What is known is that each topological invariant is connected to a family of physical phenomena. In quantum Hall materials, the transverse (Hall) conductance is an experimentally quantized invariant, corresponding in the integer quantum Hall case to the Chern number~\cite{TKNN1982,Hatsugai1993}\footnote{In the interacting case, the Hall conductance is given by a related many body Chern number and the degeneracy of the ground state. See~\cite{Niu85}.}. Explorations of synthetic quantum matter composed of ultracold atoms has resulted in new experimental observables connected to the Berry curvature~\cite{Jotzu:2014aa,Duca2015,Flaschner2015,li2016bloch}, the anomalous velocity~\cite{Aidelsburger2015}, and quantized charge transport in a Thouless pump~\cite{Thouless1983Pump,Nakajima2016,Lohse2016}; amazingly, all of these observations relate directly back to the Chern number, and each teaches us something different about its fundamental character in determining material properties. Meanwhile, understanding the physical significance of the mean orbital spin and central charge has remained challenging because the transport coefficients they impact are notoriously difficult to measure~\cite{banerjee2017observed}.

Observing phenomena associated with new topological invariants is as provocative as it is useful; such manifestations provide deep insight into the significance of otherwise opaque quantum numbers, and are real-world tools that characterize topological matter. Photonic topological materials offer especially promising routes to new experimental probes of topological invariants ~\cite{mittal2016measurement,ningyuan2015time,wang2009observation,owens2017quarter,rechtsman2013photonic,roushan2017chiral}, as they  offer the time-, energy-, position-, and momentum-resolved control available in cold-atom experiments ~\cite{Jotzu:2014aa,Duca2015,Flaschner2015,li2016bloch,Aidelsburger2015,Nakajima2016,Lohse2016,Tai2017,stuhl2015visualizing}, plus spectroscopic tools unique to electromagnetic systems~\cite{ma2016hamiltonian,mittal2016measurement,lim2017electrically}. 

In a prior work~\cite{schine2016synthetic} we demonstrated photonic Landau levels in curved space; this platform provides us new tools including (1) \emph{spatially-arbitrary excitation via holographic beam shaping}, and (2) a \emph{conical singularity of spatial curvature} that perturbs the Landau levels. In this work, we introduce (3) \emph{complex-valued tunneling spectroscopy using holographic reconstruction of the system response} to access topological invariants through spatially localized observables: harnessing a holographic reconstruction of the band projector, we measure the Chern number~\cite{kitaev2003fault,ma2016hamiltonian}; using the conical defect in the photonic Landau level, we measure the mean orbital spin and the central charge through the ``gravitational response'': the amount of density build-up, and its structure, at a singularity of spatial curvature.

We begin with a brief description of the local character of these topological invariants, connecting them to new observables. We then describe our measurement of the Chern number via a quantized bulk chiral phase response, and measurements of the mean orbital spin and chiral central charge from precision measurements of density oscillations near singularities of spatial curvature and magnetic flux. Finally, we conclude with a brief discussion of extensions of this work to interacting quantum Hall materials.

\begin{figure*}
\includegraphics[width=\textwidth]{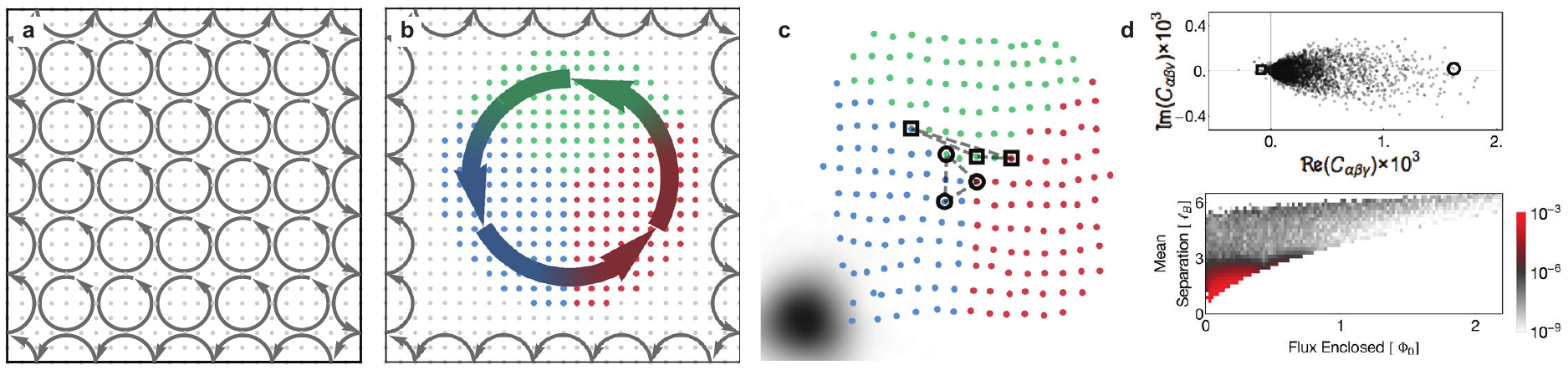}
\caption{\textbf{Chern Number Measurement in Real Space. (a)} A 2D system with a perpendicular applied magnetic field forms a bulk insulator because particles far from system edges undergo cyclotron orbits rather than linear motion. \cmmnt{: consequently application of a longitudinal field results in \emph{transverse} motion.} Associated with these bulk orbits are counter-orbiting (``skipping'') edge-trajectories. Existence of these topologically-protected 1D chiral edge channels is often the simplest-to-detect signature of a topological bulk. \textbf{(b)} Direct measurement of bulk topology requires a disorder-insensitive probe of bulk chiral non-reciprocity. We split the bulk into three adjacent but otherwise arbitrary regions and sum, for all sets of three points $\alpha,\beta,\gamma$ with one selected from each region, the non-reciprocity of the transmission amplitude $\alpha\rightarrow\beta\rightarrow\gamma$ vs $\gamma\rightarrow\beta\rightarrow\alpha$. \textbf{(c)} To employ this approach to measure the Chern invariant, we inject a TEM$_{00}$ (bottom left) magnetically displaced to points $\alpha,\beta,\gamma$ spaced by less than a magnetic length (in the first quadrant, to avoid the cone tip). \textbf{(d)} All $\sim$ 285,000 terms, $C_{\alpha \beta \gamma}$, from \textbf{c} are plotted (top), the sum of which provides a single Chern number measurement of $\mathcal{C}=1.01+0.01i$. Triples that enclose more magnetic flux and have separations of $\sim$ one magnetic length (two sites) give the largest contributions (example circled in \textbf{c} \& \textbf{d}), while triples that are far apart or do not enclose much flux provide a small contribution (example boxed in \textbf{c} \& \textbf{d}). This behavior is confirmed quantitatively (bottom) by plotting the mean contribution to $\mathcal{C}$ versus enclosed flux and mean separation of triples. The presence of non-zero imaginary components of $C_{\alpha\beta\gamma}$ reflects imperfect alignment of injection and measurement grids and cancel in $\mathcal{C}$. \label{Figure:MeasurementScheme}}
\end{figure*}
\begin{figure}[h!]
\subfloat{\includegraphics[width=1.0\columnwidth]{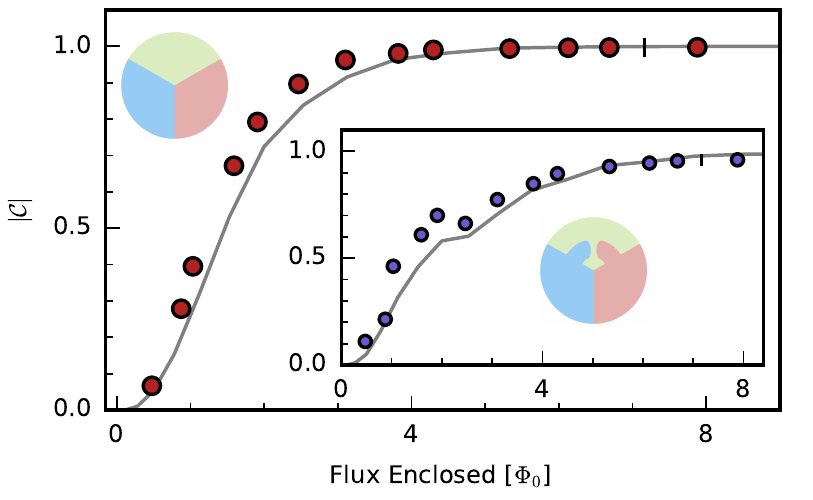}}\vspace{-1ex}\\
\caption{\textbf{Measuring the Chern Number.} The sole length scale for physics in the lowest Landau level is the magnetic length, $l_B$, so it is reasonable to expect that the Chern number will converge once the measurement area extends beyond this scale. Indeed, as the sum in Eqn. \ref{eqn:RealSpaceChern} is taken out to larger radii on a grid as in Fig. \ref{Figure:MeasurementScheme}, the measured Chern number data (red points) rapidly converges to one, in agreement with first-principles theory (gray curve) with no adjustable parameters. Errorbars are calculated from the standard deviation from 10 repetitions of the experiment and are all smaller than the points; a typical $1\sigma$ errorbar of $\pm0.02$ is plotted in place of the penultimate point. \textbf{(inset)} The Chern number is invariant to distortions of the boundaries between the three summation regions, even when the three regions approach each other at a second location.\label{Figure:ChernNumber}}
\end{figure}

\section*{New Probes of Topology}

While the Chern number $\mathcal{C}$ is traditionally defined as an integral over the Brillioun zone~\cite{TKNN1982}, the ``bulk boundary correspondence'' connects a non-zero $\mathcal{C}$ to robust chiral edge channels that extend around the boundary of the material~\cite{Hasan2010}; indeed the presence of these channels is often taken as \emph{proof} that the bulk is topological~\cite{wang2009observation, hafezi2013imaging, rechtsman2013photonic, ningyuan2015time}. Accordingly, a conceptually simple local measure of the bulk Chern number results from cutting the system down to a patch a few magnetic lengths across and surrounding it with vacuum. The number and chirality of these edge modes directly reflects the Chern number.

In practice, it is challenging to cut the system; Kitaev proposed a recipe to extract equivalent information from triple-products of spatial projectors onto a spectrally isolated band~\cite{kitaev2006anyons}. We implement this approach using spatially-resolved complex-valued tunneling spectroscopy of a patch within the bulk of the lowest Landau level, \cmmnt{to measure a non-reciprocal circulation of particles directly related to} thereby measuring a non-zero Chern number~\cite{ma2016hamiltonian}.

Two additional topological invariants appear in quantum Hall physics: the mean orbital spin $\bar{s}$ is a bulk invariant quantifying a particle's magnetic-like coupling to curvature \cmmnt{average orbital angular momentum of particles} and is related to the Hall viscosity and Wen-Zee shift; the chiral central charge, $c$, also known as the gravitational anomaly, is equal to the total number of edge modes (neutral and charged) in integer quantum Hall and Laughlin states and gives rise to the thermal Hall conductance~\cite{KaneFisher97,ReadGreen00,can2016emergent} (see Fig. \ref{Figure:Observables}). To date, these invariants have been understood in terms of gedanken experiments requiring topological gymnastics, and measured through their connection to exotic transport coefficients. We are able to access them because they govern the coupling of several local observables, namely particle- and angular-momentum- densities, to spatial curvature.

More formally, the bulk of any quantum Hall system may be described by the generic low-energy effective action $W(B,R) = f(B,R;\nu,\bar{s},c)$ ~\cite{Abanov14,Gromov15}, where $B(x,y)$ is the magnetic field and $R(x,y)$ is the spatial (Ricci) curvature. $\nu\frac{e^2}{h} = \sigma_H$ is the Hall conductance, which specifies the current induced perpendicular to an applied electric field and is precisely quantized in the famous plateaus of the integer and fractional quantum Hall effects; for integer quantum Hall physics, $\nu$ is equal to the Chern number $\mathcal{C}$. The mean orbital spin and central charge complete the triplet of topological invariants that appear in quantum Hall systems. These five quantities fully specify the effective action, whose derivatives are physical observables such as densities and transport coefficients.

From this effective action, it can be shown that the curvature localized at a cone tip produces a localized density response that depends sensitively upon both the mean orbital spin~\cite{Abanov14} and central charge (see SI ~\ref{SI:dM2toC}).

\section*{Electromagnetic Response}
%
To extract the Chern number we measure a quantized bulk chiral response~\cite{kitaev2006anyons,mitchell2018amorphous}(Fig. \ref{Figure:Observables}f). Particles inhabiting multiple Landau levels display cyclotron orbits creating a bulk circulating current. While particles in a single Landau level do not undergo cyclotron orbits, they still accrue a chiral (Aharanov-Bohm) phase when forced to travel in a closed path (Fig. \ref{Figure:MeasurementScheme}a). While this chiral phase is not apparent from the momentum-space definition of the Chern number~\cite{TKNN1982,Kohmoto85}, it is highlighted by an alternate expression~\cite{kitaev2006anyons}:
\begin{equation}
C^\mu = 12 \pi i \sum_{\alpha \in A,\beta \in B,\gamma \in C} P^\mu_{\alpha,\beta}P^\mu_{\beta,\gamma}P^\mu_{\gamma,\alpha}-P^\mu_{\alpha,\gamma}P^\mu_{\gamma,\beta}P^\mu_{\beta,\alpha}
\label{eqn:RealSpaceChern}
\end{equation}
where the area probed is split spatially into thirds labeled A, B, and C, as shown in Fig. \ref{Figure:MeasurementScheme}b and the band projector $P_{\alpha,\beta}^\mu = \left \langle \textbf{x}_\beta \right| \left \lbrack \sum_{j \in \mu} |j\rangle \langle j | \right \rbrack \left| \textbf{x}_\alpha \right \rangle$ maps eigenstates $|j\rangle$ residing in band $\mu$ to themselves and all other eigenstates to zero. Intuitively, this means injecting a tiny probe (transverse size $\ll l_B$) ~\cite{brouder2007exponential}\footnote{\protect{The Kitaev expression employs band projectors rather than Wannier orbitals because bands with non-zero Chern number do not possess exponentially-localized Wannier orbitals~\cite{brouder2007exponential}.}} $\left | \textbf{x}_\alpha \right\rangle$ at some desired location $\textbf{x}_\alpha = (x_\alpha,y_\alpha)$ and energy-integrating the resulting complex cavity response (leakage field) at another location $\textbf{x}_\beta = (x_\beta,y_\beta)$ across the band/Landau level~\cite{ma2016hamiltonian} (see Fig. \ref{Figure:Setup}c and SI \ref{SI:BandProjector}); for a Landau level/Chern band, this response is exponentially localized with a characteristic scale $l_B$/magnetic unit cell respectively.

We may then assemble triple products of these complex responses into a measurement of $\mathcal{C}^\mu$ from what are essentially chirality measurements: for any triplet of points $(\textbf{x}_\alpha,\textbf{x}_\beta,\textbf{x}_\gamma)$, the first term $P^\mu_{\alpha,\beta}P^\mu_{\beta,\gamma}P^\mu_{\gamma,\alpha}$ measures particle current in a trajectory with one handedness, $\textbf{x}_\alpha \rightarrow \textbf{x}_\beta \rightarrow \textbf{x}_\gamma \rightarrow \textbf{x}_\alpha$, while the second term measures the reverse trajectory. In magnitude, the currents are equal; however, due to the vector potential providing an Aharanov-Bohm phase for particles traversing in a closed loop, their phases are opposite. Each term in the sum is then the net non-reciprocity for that trajectory, and summing over all possible trajectories provides the Chern number (Fig. \ref{Figure:MeasurementScheme}b-d). 

We experimentally implement this protocol using a digital micromirror device (DMD) to excite each $\left| \textbf{x}_\alpha \right \rangle$ on a chosen grid \cmmnt{ of $\textbf{x}_\alpha \kern 0.02em ^s$} (Fig. \ref{Figure:Setup}a). Holographic reconstruction of the transmitted resonator electric field (Fig. \ref{Figure:Setup}a,b) while sweeping the excitation laser frequency across the Landau level then provides matrix elements of the band projector $P_{\alpha,\beta}^\mu$ from $\textbf{x}_\alpha$ to all $\textbf{x}_\beta$ (Fig. \ref{Figure:Setup}c). We obtain all projector matrix elements are obtained by iterating the excitation location over all points on the chosen grid, and the Chern number is then computed via Eqn. (\ref{eqn:RealSpaceChern}).



The terms in the sum of Eqn. (\ref{eqn:RealSpaceChern}) fall off rapidly as the points $\textbf{x}_\alpha$, $\textbf{x}_\beta$, and $\textbf{x}_\gamma$ stray from one another since $P^\mu_{\alpha,\beta}\propto e^{-|\textbf{x}_\alpha-\textbf{x}_\beta|^2 / 4l_B^2}$, so the dominant contributions to the sum come from trios of points near the meeting point(s) of the three sectors: the contribution from terms that contain any point several magnetic lengths away from the center is negligible. Accordingly, the sum in Eqn (\ref{eqn:RealSpaceChern}) can be easily truncated, as is apparent in Fig. \ref{Figure:ChernNumber}, where the Chern number is evaluated as the radius of the circular summation region is increased. Beyond a total enclosed flux of $\sim 4$ $\Phi_0$, the Chern number saturates to $\mathcal{C} = 1.00(2)$. Eqn. (\ref{eqn:RealSpaceChern}) may thus be considered a spatially localized definition of the Chern number, in the sense that an independent measurement may be made by choosing a different center of the sectors.

The quantity so-measured is \emph{indeed} an invariant, highly robust to imperfections in both the the Landau level and the measurement apparatus: nanoscopic mirror imperfections give rise to a disorder potential that weakly couples modes within the Landau level, and the excitation locations deviate from a perfect grid by $\sim 20\%$ (Fig. \ref{Figure:MeasurementScheme}c), yet the Chern number converges smoothly to $1$ (Fig. \ref{Figure:ChernNumber}). In Fig. \ref{Figure:ChernNumber}, inset, we intentionally distort the summation regions to produce a second, off-center location where all three regions approach each other (as this is where the triple-product of projectors may be largest). While the summation region must now fully enclose this new ``defect'' before the sum converges, the Chern number remains invariant to this distortion.



\begin{figure*}
\includegraphics[width=\textwidth]{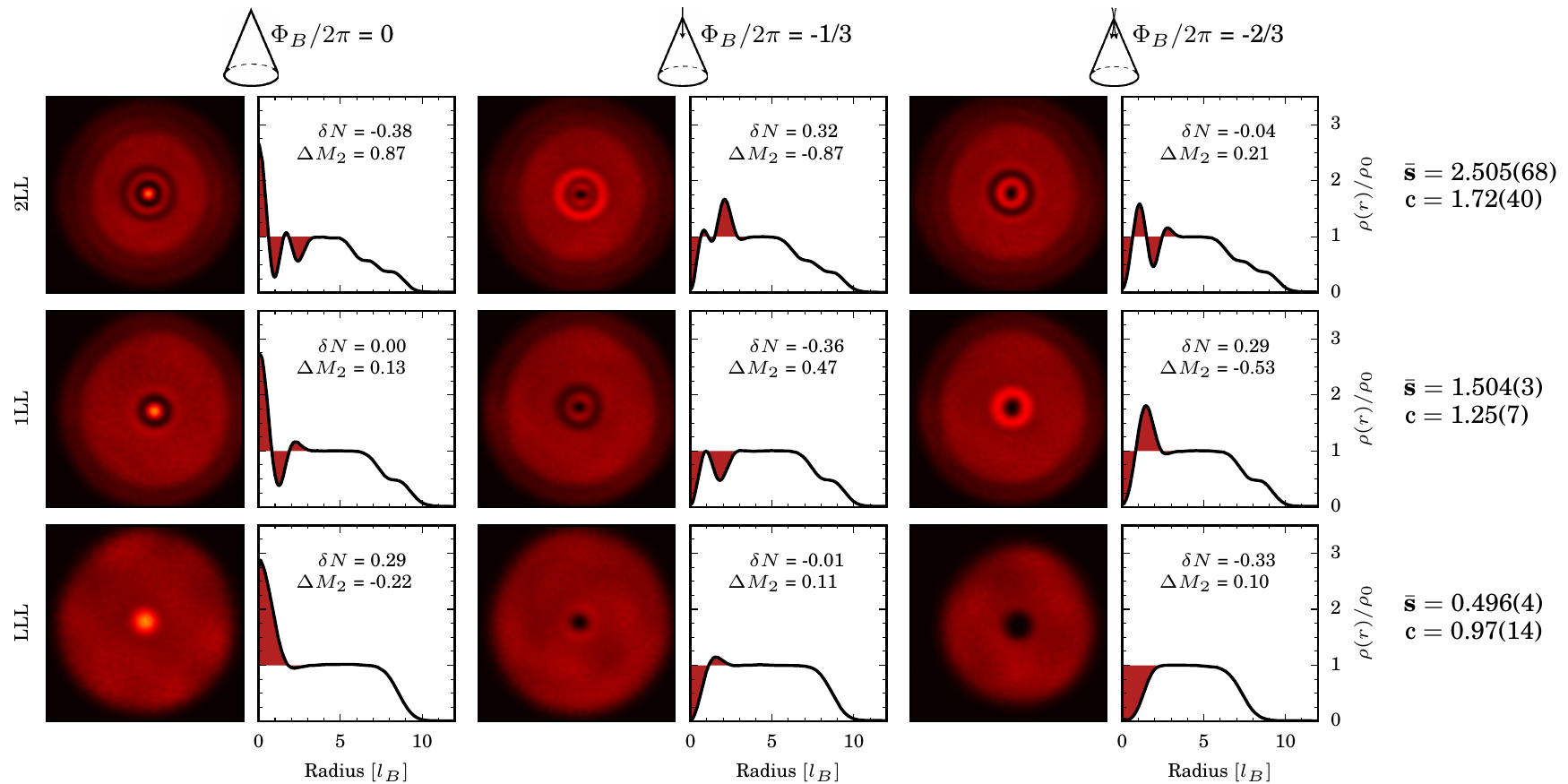}
\caption{\textbf{Response to Manifold Curvature.} A quantum Hall fluid's response to spatial curvature is governed by two additional topological invariants, the mean orbital spin and central charge. Injecting, imaging, and summing several ($\sim$ 10) single particle states provides high-precision measurements of the local state density in nine different Landau levels (red images), from which azimuthally averaged radial profiles are extracted (black curves). These data display the characteristic local density oscillations at the cone tip; a region of uniform density at larger radii; and a smooth decrease to zero due to the finite number of states measured. The uniform density region is averaged over all levels to define a background density  from which excess local state density (red filling) may be defined. For each Landau level, the excess local state density near the cone tip is integrated to measure the total excess state number, $\delta N$, from which the mean orbital spin, $\bar{s}$, is extracted. The shifted second moment, $\Delta M_2$, (see text) is also computed, which then provides a higher-precision measurement of the mean orbital spin as well as the central charge, $c$. We benchmark the connection between these topological invariants and the local density oscillations by performing the same experiment in nine situations: the lowest (LLL), first excited (1LL), and second excited (2LL) Landau levels on a cone with three possible values of magnetic flux threading its apex (left to right, flux specified at top). The errorbars in $\bar{s}$ and $c$ arise from averaging over flux threading and systematic uncertainty in the upper bound of the integration region for moment analyses. \label{Figure:ldos}}
\end{figure*}

\section*{Gravitational Response}
%

The response of a quantum Hall fluid to manifold curvature is controlled by two topological invariants, the mean orbital spin and the central charge, specific to the particular Hall state under consideration. Our platform, consisting of Landau levels on a cone with additional flux of $\Phi_B= -2\pi a/3$, $a = 0,1,$ or $2$ threaded through the tip, provides an idealized source of manifold curvature localized precisely at the cone tip. In what follows, we connect variations in the Local spatial Density of States (LDOS) at this curvature singularity directly to the mean orbital spin and central charge (see SI \ref{SI:dM2toC}):

In flat space the LDOS of a Hall fluid is uniform, providing few signatures of the fluid's properties; in curved space, however, the LDOS displays oscillations about its flat-space background that depend on the local curvature and threaded flux, $a$ \cmmnt{, and the system's topological invariants}: the excess particle number localized to the cone tip, defined as the spatial integral of the excess density there, directly reflects the mean orbital spin (Fig. \ref{Figure:Observables}g), while the width of this excess particle density reflects the orbital angular momentum attached to the curvature singularity, and thus the central charge (Fig. \ref{Figure:Observables}h). Technical improvements in the apparatus since prior lowest Landau level LDOS experiments of~\cite{schine2016synthetic} (see SI \ref{SI:ApparatusImprovements}) enable undistorted, high-precision access to these LDOS oscillations, thereby extending measurements of the mean orbital spin to excited Landau levels where the invariant takes on new values, expected to obey $\bar{s}_n = n+\frac{1}{2}$, where $n = 0,1,2,...$ specifies the lowest, first excited, and second excited Landau levels. This further provides a new and independent probe of the mean orbital spin and, most importantly, permits measurements verifying that the central charge, $c = 1$ in all Landau levels~\cite{Wen90,WenZee92,Abanov14}.

In Fig. \ref{Figure:ldos}, we present the LDOS of lowest, first excited, and second excited Landau levels on three cones differentiated by effective magnetic flux threading the cone tip, following the same procedure as in~\cite{schine2016synthetic}. Near the cone tip, we observe characteristic oscillations LDOS radial profile $\rho(r)$, which settles to a uniform background level by $r \sim 4 l_B$. At large radii, the LDOS drops to zero only because a finite number of single particle states were included, the number being limited by the size of the DMD used for mode injection. The background level is equal for all nine LDOS measurements (see SI \ref{SI:rad_profiles}), and their average is used to define the background local state density $\rho_0$ for all measurements. We then compute the total excess particle number, $\delta N = \int (\rho(\textbf{r})-\rho_0) \Diff2 \mathbf{r}$, and a measure of the excess density's width, the shifted second moment, $\Delta M_2 = \int (\rho(\textbf{r})-\rho_0)\left(r^2/2-(2n+1)\right) \Diff2 \mathbf{r}$. These quantities then provide the mean orbital spin and central charge (as the primary theoretical result of this article, see SI \ref{SI:dM2toC} and \ref{SI:dNtoSbar}) 

From the excess particle number, we measure the mean orbital spin and average the result over flux threading in the lowest three Landau levels, finding $\bar{s} = \{0.47(3), 1.47(3), 2.45(3)\}$ for $n = 0,1,$ and $2$, respectively. We can also use measurements of $\Delta M_2$ to extract the mean orbital spin, as the linear component of the dependence of $\Delta M_2$ on the flux $a$ is exactly $(\bar{s}-n)a$ (See SI \ref{SI:dM2toC}). This provides a significantly more precise determination of $\bar{s} = \{0.496(4), 1.504(3), 2.505(68)\}$. We then use these measurements of $\bar{s}$ along with measurements of $\Delta M_2$ to calculate the central charge in each Landau level, finding $c = \{1.0(1), 1.3(1), 1.7(4)\}$. While the precision of the central charge measurement drops in higher Landau levels due to finite field of view and increased sensitivity to error in the mean orbital spin, all mean orbital spin measurements and the lowest Landau level central charge measurement are in agreement with theoretical expectations for the integer quantum Hall fluid.

\section*{Outlook}
In this work we have developed and measured local observables that characterize bulk invariants of topological materials. Our approach elucidates the physical significance of these invariants and relaxes the non-physical sensitivity of the standard definitions to experimental imperfections like disorder. Indeed, the TKNN formulation of the Chern number assumes discrete translational symmetry~\cite{TKNN1982}.


While Hall conductance, mean orbital spin, and central charge do not fully characterize a \emph{generic} quantum Hall state, they often provide sufficient information to distinguish between candidate phases in the lab. In the case of the electronic $\nu = 5/2$ fractional quantum Hall plateau, a measurement of \emph{either} the mean orbital spin or central charge would suffice to choose amongst the more-than nine candidate states~\cite{Banerjee17}. The photonic analog is bosonic, so similar physics is expected at $\nu = 1$, permitting the exploration of and differentiation between Pfaffian and parafermion states~\cite{Cooper01,Regnault03}.

Exploring such interacting topological phases of photons~\cite{hafezi2013non} will require combining Landau levels of light in twisted resonators~\cite{schine2016synthetic} with Rydberg-mediated interactions between photons~\cite{jia2017strongly,peyronel2012quantum}. Such phases may be assembled particle-by-particle~\cite{CarusottoFQH2012,grusdt2014topological,dutta2017coherent} or by dissipative stabilization~\cite{PhysRevX.4.031039,lebreuilly2017stabilizing}; in either case, measurement of the gravitational response will be an essential tool for characterization of the resulting topological phase~\cite{Wu2017FQHEcones}. Furthermore, extension of the bulk circulation measurement to the strongly interacting regime has the potential to permit direct observation of anyon braiding statistics.

\section*{Acknowledgements}
We would like to thank Charles Kane and Michael Levin for fruitful conversations. This work was supported by DOE grant DE-SC0010267 for apparatus construction/data collection and MURI grant FA9550-16-1-0323 for analysis.

\section*{Author Contributions}
N.S., M.C., and J.S. designed and built the experiment. N.S. and M.C. collected and analyzed the data. Development of the theory concerning $\bar{s}$ and $c$ was performed by T.C. and A.G. All authors contributed to the manuscript.

\section*{Author Information}
The authors declare no competing financial interests. Correspondence and requests for materials should be addressed to J.S. (simonjon@uchicago.edu)

\bibliographystyle{unsrt}
\bibliography{theBib.bib}

\incltocpage

\clearpage

\tableofcontents
\appendix

\section{Methods}
\label{SI:Methods}
%

The experimental resonator consists of four 100 mm radius-of-curvature high reflectivity mirrors coated for both 780 nm and 1560 nm mounted in two steel structures which define a stretched-tetrahedral resonator geometry characterized by an axial length of 5.1 cm and an opening half-angle of 10$^\circ$. The two steel mounts are aligned via rods, and a micrometer stage controls the relative separation. This permits smooth length adjustment to tune the resonator to degeneracy. One mirror is mounted on a piezoelectric transducer, which permits stabilizing the cavity length via PI feedback based on a Pound-Drever-Hall error signal generated by the reflection 1560 nm light off a resonator mirror.

To excite the resonator with light with arbitrary amplitude and phase profiles, we shine 780 nm narrowband laser light onto computer generated holograms produced by a phase-corrected digital micro-mirror device (DMD), and we direct the resulting diffracted light into the resonator. We then extract the full amplitude and phase information of the transmitted resonator field by interference with a reference beam (See SI \ref{SI:EfieldReconstruction}). 

We employ the DMD as a generalized scanning tunneling microscope, enabling us to inject light with arbitrary position, momentum, or angular momentum. \cmmnt{Near the cone tip, the singularity of spatial curvature perturbs the mode structure the injection of regular, flat-space modes.} Remarkably, the holographic reconstruction technique allows us to measure the (spatially localized) band projectors even when disorder or harmonic confinement breaks the degeneracy between the modes in the Landau level; it is only necessary that the probe sweep across the band of states in the Landau level, and that the resulting resonator response be interfered with the reference beam before ``integration'' of the intensity on an camera. At each laser-frequency in the sweep, the resonator response will be a ring carrying orbital angular momentum; these rings interfere can interfere with one another, even if they arrive on the camera at different times, because the interference with the heterodyne beam converts phase to intensity, resulting in the desired localized mode (Fig. \ref{Figure:Setup}c). 


\begin{figure}[h]
\subfloat{\includegraphics[width=.45\columnwidth]{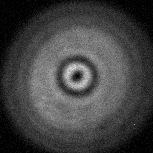}}
\subfloat{\quad}
\subfloat{\includegraphics[width=.45\columnwidth]{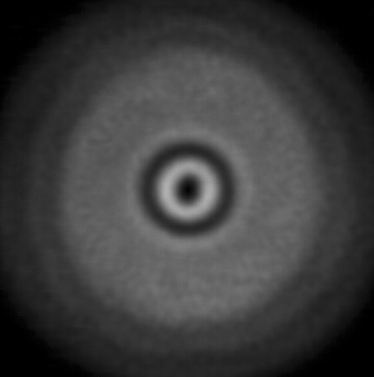}}
\caption{\textbf{Resonator Imaging Comparison.} The local density of states in the second excited Landau level with effective magnetic flux $\Phi_B/2\pi = -2/3$ threading the cone tip highlights improvements in resonator design. The previous resonator used in~\cite{schine2016synthetic} (left), displays significant diagonal astigmatism, which has been removed in the current work (right). The imaging system is also improved, and the total number of modes accessed has been increased, providing a more precise determination of the background density.\label{Figure:LDOScomparison}}
\end{figure}

\section{Apparatus Improvements}
\label{SI:ApparatusImprovements}

The current apparatus is based on that used in our previous work~\cite{schine2016synthetic}. We rebuilt the experimental resonator with a new mirrors, mirror mounts, and in a new configuration. The resonator housing is now steel rather than plastic, and the resonator length is stabilized with a Pound-Drever-Hall error signal controlling proportional-integral feedback circuitry which actuates a piezo stack glued to a mirror. This enables the precise control of the probe laser detuning from the resonator resonance necessary for the measurement of local projectors. We image the transverse plane of the resonator by collecting light transmitted through one of the mirrors. In passing through the glass substrate of a curved mirror at significant non-normal incidence, the light is defocused by an effective cylindrical lens. This appears as an artificial breaking of rotational symmetry in the resonator modes and had previously limited LDOS measurements. In particular, it made measurements of the second moment impossible, since unlike the integrated excess density, the second moment is not invariant to astigmatic distortion. Both increasing the mirror radii of curvature from two at 25 mm and two at 50 mm to all four at 100 mm and reducing the non-planar opening half-angle from 16$^\circ$ to 10$^\circ$ serve to reduce the effect of this aberration (see Fig. \ref{Figure:LDOScomparison}).

\begin{figure*}[t]
\includegraphics[width=.95\textwidth]{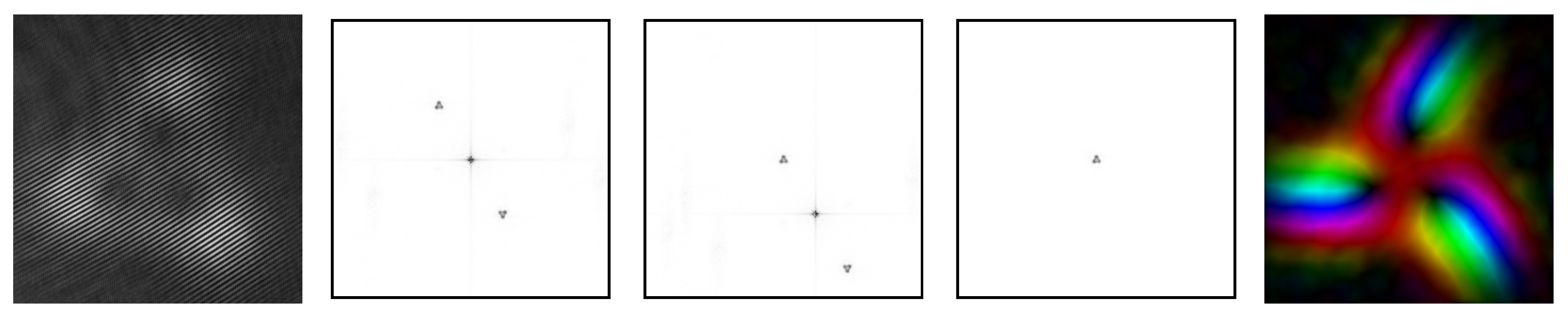}
\caption{\textbf{Holographic Electric Field Reconstruction.} A heterodyne image is obtained, (\textbf{a}), showing the cavity mode intensity with high frequency fringes superimposed. Background and cavity field images are subtracted from the heterodyne image, and a 2d spatial Fourier transform is applied, providing the momentum space image, (\textbf{b}). This reveals two copies of the unmodulated cavity mode profile shifted by plus and minus the $k$-vector of the heterodyne beam, with a large DC background in the middle. The momentum space image is then shifted, (\textbf{c}), and Gaussian masked, (\textbf{d}). Finally, an inverse Fourier transform provides the complex-valued electric field profile of the cavity mode (\textbf{e}).\label{Figure:FullEfieldReconstruction}}
\end{figure*}

\begin{figure}[t]
\subfloat{\includegraphics[width=0.6\columnwidth]{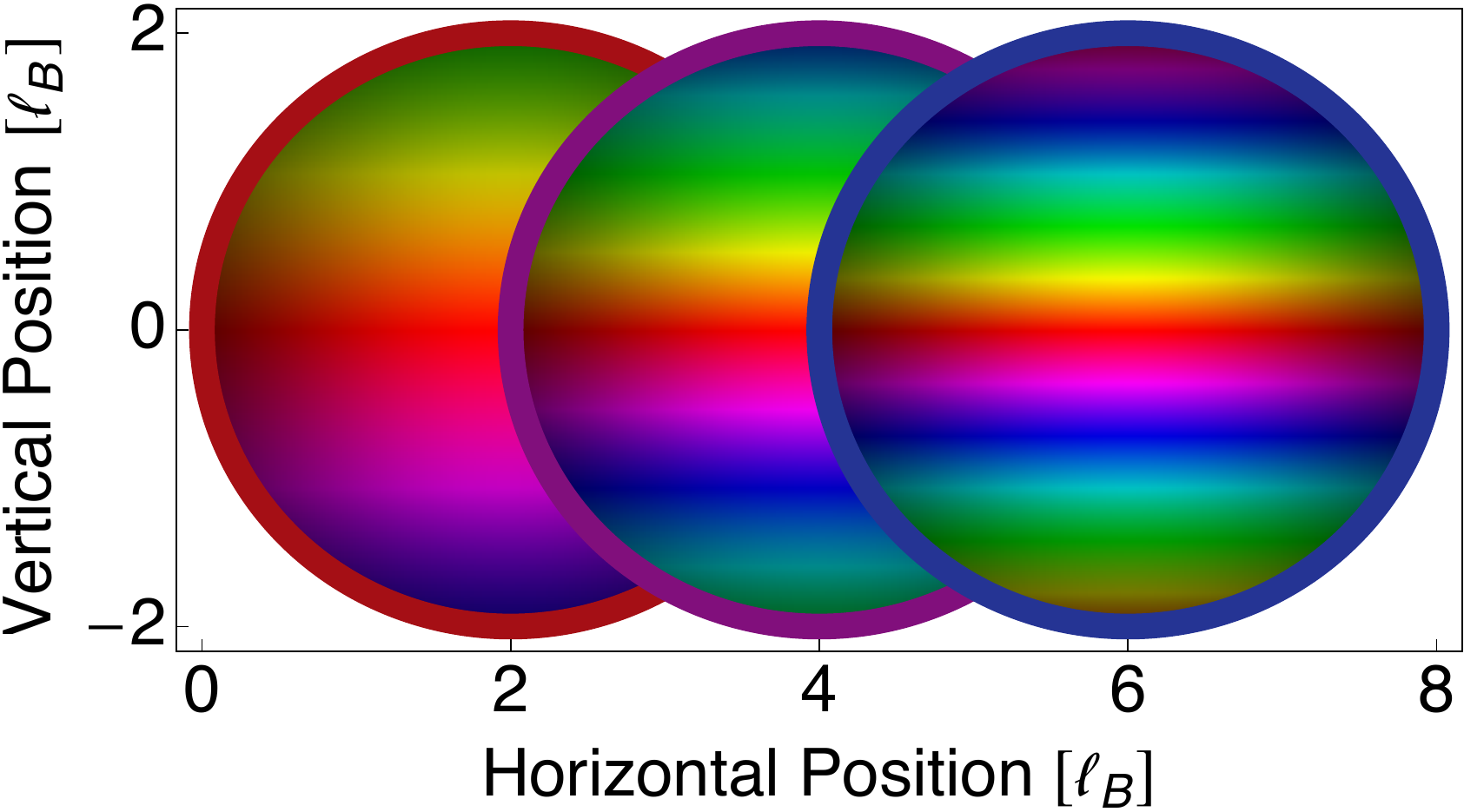}}\\
\subfloat{\includegraphics[width=.82\columnwidth]{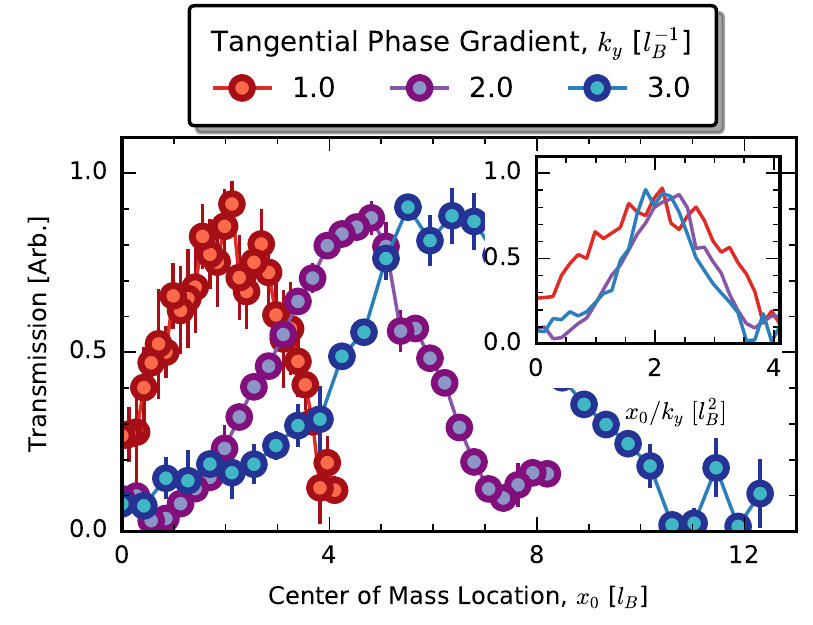}}\\
\subfloat{\includegraphics[width=.82\columnwidth]{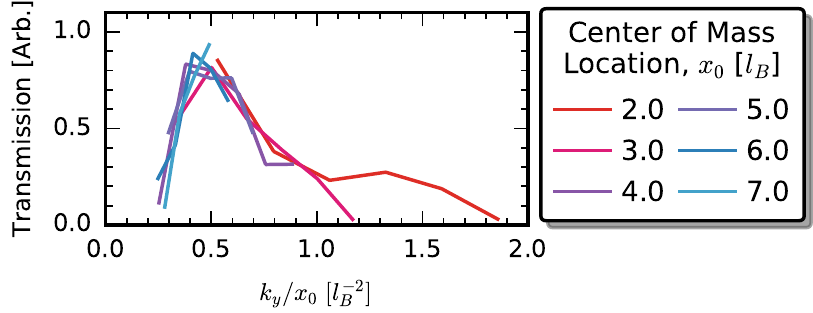}}
\caption{\textbf{Chiral-Only Resonator Modes. (a)} Our photonic Landau level only supports modes of a particular chirality. As an injection point is displaced horizontally, this implies a phase gradient vertically which is proportional to the horizontal displacement. \textbf{(b)} We test the chirality of resonator modes by injecting light at various locations and tilts, with the total transmission taken as a measure for the overlap of the probing light with modes in the Landau level. The total transmission through the resonator is plotted as the radial displacement is increased with constant tangential phase gradient, $k_y$. When transmission is plotted versus the ratio between the tangential phase gradient and the radial displacement (inset), a distinct peak at $x_0/k_y=2 l_B^2$ is observed independent of the magnitude of the tangential phase gradient. The peaked structure of this plot indicates the chiral structure of modes supported by the resonator. Error bars represent the standard deviation of 10 identical runs of the experiment. \textbf{(c)} The same data may be re-sliced to observe transmission versus $k_y$ at constant $x_0$, corresponding to cuts orthogonal to those taken in \textbf{b}. When the transmission is plotted versus the ratio $k_y/w_0$, we again observe a distinct peak at $k_y/x_0=0.5 l_B^{-2}$ independent of the magnitude of the radial displacement. \label{Figure:TangentialPhaseGradient}}
\end{figure}

\section{Holographic Measurement of Electric Field}
\label{SI:EfieldReconstruction}

Heterodyne imaging provides a phase reference for the cavity output field, allowing the extraction of not just amplitude but also phase information. The heterodyne image appears similar to an image of just the cavity output except with the addition of high frequency fringes. Much like RF modulation spectroscopy, a low noise, high quality image of the original field may be obtained by an appropriately chosen ``demodulation'' scheme, depicted in Fig. \ref{Figure:FullEfieldReconstruction}. 

The heterodyne image is given in term of the cavity mode electric field, $E_c(\textbf{x})$, and the heterodyne field, $E_{het}(\textbf{x}) = E_h(\textbf{x}) \, e^{i \textbf{k}\cdot \textbf{x}}$, by
\begin{equation}
I_{het}(\textbf{x}) = |E_c(\textbf{x})+E_h(\textbf{x})\,e^{i \textbf{k}\cdot \textbf{x}}|^2,\nonumber
\end{equation}
where $E_h(\textbf{x})$ is a slowly varying function.
After subtracting off a heterodyne beam background image $|E_h(\textbf{x})|^2$ and a cavity mode image $|E_c(\textbf{x})|^2$, the signal is given by
\begin{equation}
I_{sig} = E_c(\textbf{x}) \,E_h^*(\textbf{x})\, e^{-i\textbf{k}\cdot\textbf{x}}+E_c^*(\textbf{x})\,E_h(\textbf{x})\, e^{i\textbf{k}\cdot\textbf{x}}.\nonumber
\end{equation}
Extracting just one component of this via Fourier space filtering then yields 
\begin{align}
I_{demod}&=E_c(\textbf{x}) \,E_h^*(\textbf{x})\nonumber\\
&\propto E_c(\textbf{x}),
\end{align}
where the final proportionality follows assuming the heterodyne beam was a clean plane wave with negligible variation across the cavity mode. That the demodulated signal is proportional to the square root of the intensity of the heterodyne beam indicates the suitability of this technique to the measurement of very low cavity field amplitudes, requiring in that case a camera with high dynamic range. 

It is also worth noting that the subtraction of the individual heterodyne and cavity mode images is, in practice, often not necessary. To improve spatial resolution of the phase measurement, it is advantageous to make the heterodyne beam produce short wavelength fringes with a period approaching $\sqrt{2}a$, where $a$ is the pixel size and the direction of the fringes is at 45$^\circ$ to the pixel axes. This also ensures that the modulated electric field is maximally separated from slowly spatially varying ``DC" backgrounds. As long as the cavity modes imaged onto the camera cover many pixels, the cavity mode background $|E_c(\textbf{x})|^2$ will appear as a slowly varying DC background. By assumption, the same holds of the heterodyne beam, so both the cavity mode and heterodyne backgrounds will be removed by the Fourier space masking (Fig. \ref{Figure:FullEfieldReconstruction}c,d). 

\section{Connecting The Band-Projector To The Cavity Response Of A Swept Laser And Its Subsequent Normalization}
\label{SI:BandProjector}

The definition of the projector onto a band $\mu$ is $P^\mu\equiv \sum_{j \,in\, \mu}|j\rangle \langle j|$ for $|j\rangle$ in an orthonormal basis. From this it follows that $(P^\mu)^2=P^\mu$.

We can then define matrix elements of the projector between localized modes injecting at $|\textbf{x}\rangle$ and measuring at $|\textbf{y}\rangle$ as $P^\mu(\textbf{x},\textbf{y}) \equiv \langle\textbf{y}|P^\mu|\textbf{x}\rangle$. From this definition and $(P^\mu)^2=P^\mu$, it follows that 
\begin{equation}
P^\mu(\textbf{x},\textbf{x}) = \sum_{\textbf{y}}|P^\mu(\textbf{x},\textbf{y})|^2.
\label{Eqn:ProjNormCondition}
\end{equation}
This expression forms the normalization criterion for measured matrix elements of the projector. 

\droptocpage
\subsection*{Measuring the projector from the cavity response}
We wish to measure the projector onto a Landau level by measuring some response function of the cavity to some probe. Here, we show that matrix element of the projector between two points $\textbf{x}$ and $\textbf{y}$ is equal to the value of the electric field at $\textbf{y}$ of the cavity response to an excitation at $\textbf{x}$.

We suppose that the cavity has a Hamiltonian, $H$, with or without interactions, and which has complex eigenvalues $\varepsilon_j\equiv\omega_j+\frac{i}{2}\Gamma_j$. Following~\cite{ma2016hamiltonian}, we perform first order non-Hermitian perturbation theory to find the response of the cavity to some excitation. It is worth noting that this perturbative approach is exact for linear systems such as the one described in the main text.

A weak probe of frequency $\omega$ exciting a location $\textbf{x}$ is described by an operator $\tilde{V}_{\textbf{x}}$ applied to the vacuum state $|0\rangle$. The transmitted cavity field is then given by
\begin{equation}
|\psi\rangle = \frac{1}{\textbf{1}\omega-H}\tilde{V}_{\textbf{x}}|0\rangle. \nonumber
\end{equation}

The value of this field at $\textbf{y}$ is then $\langle\textbf{y}|\psi\rangle$. When we integrate across the band of interest, the response $\phi$ is then
\begin{equation}
\phi = \langle \textbf{y}|\sum_j\int d\omega\frac{|j\rangle\langle j|}{\omega-\varepsilon_j}|\textbf{x}\rangle. \nonumber
\end{equation}
We now evaluate the integral. In practice we do not integrate over all frequencies, so we specify the limits of integration to cover a range centered at some frequency $\omega_0$ with range $\Omega$. Since this integral must be taken for each term in the sum, we therefore write
\begin{align}
\phi &= \sum_j \langle \textbf{y}|j\rangle\langle j|\textbf{x}\rangle \int_{\omega_j+\delta_j-\Omega/2}^{\omega_j+\delta_j+\Omega/2}\frac{d\omega}{\omega-(\omega_j+\frac{i}{2}\Gamma_j)} \nonumber\\
&= \sum_j \langle \textbf{y}|j\rangle\langle j|\textbf{x}\rangle \log\left(\frac{\delta_j-\frac{i}{2}\Gamma_j+\Omega/2}{\delta_j-\frac{i}{2}\Gamma_j-\Omega/2}\right) \nonumber\\
&= \sum_j \langle \textbf{y}|j\rangle\langle j|\textbf{x}\rangle \left(i \pi + \frac{\delta_j-\frac{i}{2}\Gamma_j}{\Omega/4}+\mathcal{O}\left(\frac{1}{\Omega^2}\right) \right)
\end{align}
where $\delta_j \equiv \omega_0-\omega_j$ are the individual eigenstates' detunings from $\omega_0$. In the last step we have performed a Taylor expansion in $\frac{\delta_j-\frac{i}{2}\Gamma_j}{\Omega/2}$ since we assume we sweep over a range large compared to the the individual resonances' widths and detunings from $\omega_0$. The zeroth order term in $\phi$ directly provides the projector $P^\mu(\textbf{x},\textbf{y})$ so long as the integration completely covers the band $\mu$ while avoiding all other states, while the first order term allows to estimate our error from finite and off-center integration over frequency.

\subsection*{Normalization of the measured projector}
A given heterodyne image provides the electric field everywhere in the transverse plane of the cavity given a particular input location. Taking an entire scan over input locations can take $\sim 10$ minutes, so there could be significant drifts in the experimental apparatus between two images taken at the beginning and end of a run. This is particularly important to consider since each term in the Chern number sum compares the response of the cavity to three often well separated injection locations. As such we consider measurements of the projector matrix elements, $p^\mu(\textbf{x},\textbf{y})$, which contain imperfections that are constant within an image, but vary between images: $P^\mu(\textbf{x},\textbf{y})=p^\mu(\textbf{x},\textbf{y})\gamma^\mu(\textbf{x})$. Imposing Eqn. (\ref{Eqn:ProjNormCondition}) then determines the normalization factor.
\begin{equation}
\gamma^\mu(\textbf{x}) = \frac{(p^\mu(\textbf{x},\textbf{x}))^*}{\sum_\textbf{y}|p^\mu(\textbf{x},\textbf{y})|^2}
\end{equation}
where $z^*$ indicates the complex conjugate of $z$.

\incltocpage
\section{Comparison to Fully Degenerate Cavities}
Our measurement of the Chern number works because the number of modes in the lowest Landau level, like in a discrete system, is proportional to the area, and there is a smallest feature size (set by the magnetic length) which defines a ``unit cell". This latter fact is reflected in the unusual commutator $[\hat{X},\hat{Y}]=il_B^2$, where $\hat{X}$ and $\hat{Y}$ are the coordinates of the center of semi-classical cyclotron orbits/Landau-level projected coordinates~\cite{haldane2011geometrical,bellissard1994noncommutative}. 

A similar Chern number measurement could be made in a resonator with all transverse modes degenerate; the cavity response would then always be determined by the size of the DMD-generated probe light. However, this would be incidental as there would be no smallest feature size and non-circulating and counter-circulating modes would be supported: the measured Chern number would sensitively-depend upon the fidelity of the DMD and would not be quantized. The present work explores a nearly-degenerate multimode resonator where the only modes in a band of near-degenerate states are precisely those comprising a Landau level, so the resonator does not support non-circulating or counter-circulating excitations (at the same energy). In fact, we need not assume anything about the mode structure of our resonator to perform the Chern number measurement: we can probe with a very tightly focused beam and find the frequency bands at which the resonator transmits. So long as the probe is spatially small enough (so that the phase profile mismatch between the probe field and cavity modes is irrelevant), it will excite chiral resonator modes from which the non-zero Chern number may be computed. In practice, making a very small probe size results in a small transmitted signal, but making the probe larger then only increases the signal if the phase profile of the probe matches the chiral modes of the cavity.

To investigate this, we excite the resonator at different locations $(x_0,0)$ with a constant rate of tangential phase increase,  $\exp{i k_y y }$, and with a variable rate of tangential phase increase at constant location. We observe clear maxima near $x_0/k_y=2 l_B^2$ (Fig. \ref{Figure:TangentialPhaseGradient}b, inset) and $k_y/x_0 = 0.5 l_B^{-2}$ (Fig. \ref{Figure:TangentialPhaseGradient}c) demonstrating that \emph{only} modes with a particular chirality are supported. This is to say that the supported cavity modes are the TEM$_{00}$ mode be \emph{magnetically} translated away from the cavity axis, $E_{x_0,y_0}(x,y) = E_{00}(x-x_0,y-y_0) \exp{\left(i(x_0y-y_0x)/2l_B^2 \right)}$. The spatial profile of the probe serves only to specify how strongly the chiral cavity modes are excited; it does not affect the measured Chern number. We have explicitly verified this insensitivity of the Chern number by operating at $k_y/x_0 = 0.75l_B^{-2}$, away from the peaks in Fig. \ref{Figure:TangentialPhaseGradient}b,c. This reduces the signal in the heterodyne images; however, the Chern number remains quantized at $C=1$.

\section{LDOS radial profiles}
\label{SI:rad_profiles}
\begin{figure}[h!]
\includegraphics[width=\columnwidth]{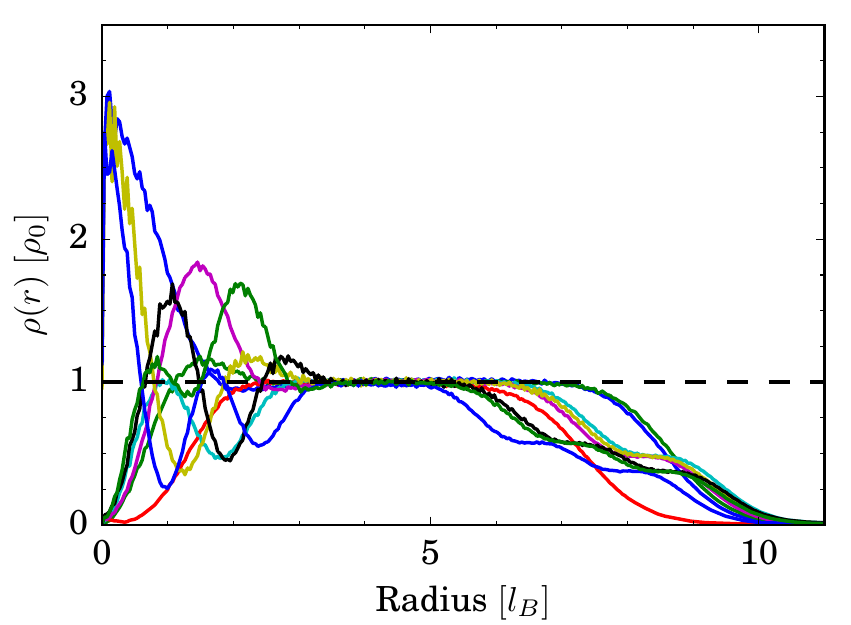}
\caption{\textbf{LDOS Radial Profiles.} All nine LDOS radial profiles are scaled uniformly and superimposed. By $\sim 4 l_B$, all profiles have settled to a consistent constant value which is averaged to define the background state density $\rho_0$.\label{Figure:ldos_rads}}
\end{figure}

In figure ~\ref{Figure:ldos_rads}, we show the angle-averaged LDOS for all three cones and all three Landau levels, superimposed. The large-radius asymptote is employed as the background density $\rho_0$.

\section{Local Measurements of Topological Quantum Numbers}
\label{SI:dM2toC}

We show how topological quantum numbers characterizing the quantum Hall effect can be measured from the density of states. We review the response of QH states in the lowest Landau level to magnetic and geometric singularities~\cite{can2016emergent}, and generalize these results to higher Landau levels. For concreteness, we consider magnetic singularities as magnetic field configurations with a delta function source

	\begin{align}
	B = B_{0} - a \phi_{0} \delta^{(2)}({\bf r})\label{B-defect}
	\end{align}
where $\phi_{0} = h/e$ is the flux quantum, and curvature singularities as geometries which have a point with a delta function curvature

\begin{align}
R = 4\pi \left( 1 - \frac{1}{s}\right) \delta^{(2)}({\bf r})\label{R-defect}	
\end{align}

The electromagnetic and geometric response of a QH state will influence both the total charge and orbital angular momentum (OAM) which accumulates at these defects. We will show how the charge and OAM are related to topological quantum numbers, and how they can be computed from the density of states. 


\subsection{Charge and OAM from Induced Action}

The charge density and the spin density follow from the variational formula 

\begin{align}
\rho & = \frac{\delta W[A, g]}{\delta A_{0}}, \quad \rho^{s}= \frac{\delta W[A,g]}{\delta \omega_{0}}\label{densities}
\end{align}
where the induced action $W[A, g]$ is a functional of the gauge potential $A_{\mu}$ and the background metric $g_{\mu \nu }$, $A_{0}$ is the time component of the gauge field (i.e. the scalar potential), and $\omega_{0}$ is the time component of the spin connection. 

The induced action is given by~\cite{Abanov14,Gromov15}
\begin{equation}
W = \frac{\nu}{4\pi}\int AdA + \frac{\nu \bar{s}}{2\pi}\int Ad\omega - \frac{c-12\nu \bar{s}^{2}}{48\pi}\int \omega d\omega.
\end{equation}
This is a functional of the vector potential $A_i$ and the background metric $g_{ij}$. The spin connection, $\omega_i$, is defined through $\nabla \times \omega = R/2$ where R is the Ricci curvature. The integrands use a compressed ``form'' notation $Ad\omega = \epsilon^{\mu\nu\rho}A_\mu\partial_\nu\omega_\rho \Diff2 x \diff t$ where $\epsilon$ is the absolutely antisymmetric tensor, and the indices $\mu$, $\nu$, and $\rho$ cover both spatial dimensions and time. 

\smallskip
{\bf Charge Density}
The charge density which follows from the the topological action is given by the well known result
\begin{align}
\rho = \frac{\nu }{2\pi} B + \frac{\nu \bar{s}}{4\pi} R	\label{density}
\end{align}
where $R$ is the scalar curvature. Integrated over a closed surface, this yields the relation between the total number of particle $N$ and the total flux $N_{\phi}$ through the surface (in units of the flux quantum ): $N = \nu N_{\phi} + \nu \bar{s} \chi$, where $\chi$ is the Euler characteristic of the surface. The second term on a sphere ($\chi = 2$) defines the shift $\mathcal{S} = 2 \bar{s}$. In the presence of a cone with a magnetic flux threading it, we can simply insert the singular fields (\ref{B-defect}\ref{R-defect}) into (\ref{density}) and integrate to find
\begin{align}
Q_{tip} = \int \left( \rho - \frac{\nu}{2\pi l^{2}}\right) dV = - \nu a + \nu \bar{s} \left( 1 - \frac{1}{s}\right)	\label{Qtip-action}
\end{align}

where $l^{2} = \hbar / (eB)$ is the magnetic length, and $\nu/(2\pi l^{2})$ is the value of the density far away from the cone tip. 
\smallskip

{\bf OAM density}

The spin density corresponds to the extensive part (scaling as $N$) of the orbital angular momentum (OAM) density, and is given by

\begin{align}
\rho^{s} = \frac{\nu \bar{s}}{2\pi} B - \frac{(c - 12 \nu \bar{s}^{2})}{48\pi } R	
\end{align}

Integrated over the surface this yields the Hall viscosity coeffiicent with a finite size correction~\cite{Gromov15}

\begin{align}
\eta^{H} = \int \frac{1}{2}\rho^{s} dV = \frac{\nu \bar{s}}{2} N_{\phi}	 - \frac{(c - 12 \nu \bar{s}^{2})}{24 } \chi.
\end{align}

In the presence of a magnetic and geometric singularity, the induced action will have additional contributions due to the singularities. Simply plugging in the singular field configurations will make the action infinite. For this reason, we need another approach to access the OAM due to the singular defects. This approach combines the microscopic definition of the OAM with the conformal block construction of FQH states. 

\subsection{OAM from moments of density}

The single-particle eigenstates of the non-interacting Hamiltonian

\begin{align}
H = \frac{1}{2m} \left(\Pi_{x}^{2} + \Pi_{y}^{2}\right), \quad 	\Pi_{i} = p_{i} - e A_{i}
\end{align}
with the fields given by (\ref{B-defect})and (\ref{R-defect}) are (in radial coordinates with magnetic length $l = 1$)

\begin{align}
\Psi_{k}^{(n)}(r, \phi) &= \frac{1}{\sqrt{\mathcal{Z}}} e^{i k \phi} r^{ |k|} e^{ - r^{2}/4} L_{n}^{|k|}(r^{2}/2)	,\label{eigenstates} \\
\mathcal{Z} & = \frac{2^{k+1} \pi \Gamma(n + k + 1)}{s \Gamma(n+1)}\\
 E &= \hbar \omega_{B} \left( n + \frac{1}{2}(|k| - k) + \frac{1}{2}\right) .\label{energy}
\end{align}

Here $n = 0, 1, ...$ labels the Landau level (LL) index, $k = s m + a$, with $m = 0, \pm 1, \pm 2, ....$, and $L_{n}^{k}(z)$ are the associated Laguerre polynomials. These wave functions are expressed in symmetric gauge to exploit the rotational symmetry of the Hamiltonian. For this reason, the orbital angular momentum commutes with the Hamiltonian and is a good quantum number. The microscopic definition of the OAM is

\begin{align}
\hat{L} = \epsilon_{ij} r_{i} \Pi_{j} + \frac{\hbar }{2} r^{2}	
\end{align}

where $\epsilon_{ij}$ is the antisymmetric symbol, and $r = \sqrt{r_{x}^{2} + r_{y}^{2}}$. 
The absolute value of the OAM in a single-particle state $\Psi_{k}^{(n)}$ which satisfies $\hat{L} \Psi_{m}^{(n)} = \hbar k \Psi_{k}^{(n)}$ can be expressed as a {\it shifted second moment} of the probability density

\begin{align}
\hbar |k| = \hbar \int \left( \frac{r^{2}}{2} - 2n -1 \right) | \Psi_{k}^{(n)}|^{2} dV	 , 
\end{align}
where the volume measure used here is $dV = \frac{2\pi}{\lambda} r dr$. This fact follows by explicit computation of the expectation value using the eigenstates (\ref{eigenstates}). 

\smallskip

\subsection{Many-particle states in the LLL and OAM of defects} 
For an $N$-particle state (interacting or not) in the lowest LL, this formula generalizes in the obvious way
\begin{align}
\langle \hat{L}_{tot} \rangle_{N} = \int \left( \frac{r^{2}}{2} - 1\right)  \langle \rho(r) \rangle  dV
\end{align}
where $\hat{L}_{tot} = \sum_{i = 1}^{N} \hat{L}_{i}$ and $\hat{L}_{i}$ acts only on the coordinate of the $i$th particle. This expression gives the total OAM of the many-particle state. From this, we have to assign an OAM due to the presence of the magnetic or geometric singularity at the origin. Here, we use an important property of QH states which states that density correlations are exponentially suppressed on scales of order magnetic length $l_B$ (which has been set to unity). This means that sufficiently far from the defect, the density will return to its mean value if the defect were not present. In the QH case, this is just $\langle \rho \rangle \to \frac{\nu}{2\pi }$, where $\nu$ is the filling fraction. Thus, the OAM of the cone tip should be captured in the moment formula 
\begin{align}
L_{tip}= \int \left( \frac{r^{2}}{2} - 1\right) \left( \langle \rho(r) \rangle  - \frac{\nu}{2\pi }\right) dV \label{OAM_LLL}
\end{align}

In Ref.~\cite{can2016emergent}, this was shown for Laughlin states to be equal to 
\begin{align}
L_{tip} = \frac{c - 12 \nu \bar{s}^{2}}{24} \left( s - \frac{1}{s}\right) + \frac{a}{2} \left( 2 \bar{s} - \frac{a}{s}\right)	\label{OAM_FQH}
\end{align}
where $c = 1$ and $\bar{s} = \frac{1}{2}\nu^{-1}$. Here, $ a(h / e)$ is the total flux threading the cone tip. The first term can be interpreted as the ``spin" of the conical defect, while the second term (in the Laughlin case) can be interpreted as the spin of a quasihole with total charge $- \nu a/s$. 

We intentionally write Eqn.(\ref{OAM_FQH}) in a form which suggests generalization to arbitrary FQH states. In fact, in~\cite{gromov2016geometric,can2017quantum}, it was shown that the angular momentum due to a cone tip is a consequence of the gravitational anomaly occurring in the CFT construction of fractional QH wave functions~\cite{ferrari2014fqhe,bradlyn2015topological}. This connection is rather natural, since the gravitational anomaly controls the behavior of the wave function under scale transformations $z \to \lambda z$, while the angular momentum is read out by considering the special case $\lambda = e^{ i \theta}$ which corresponds to pure rotations.


\subsection{Integer quantum Hall states in Higher Landau levels}

The generalization of Eqn.(\ref{OAM_LLL}) to higher Landau levels is straightforward, but somewhat subtle on a conical singularity. We begin by stating the result, and proceed to unpack it in the following section. 

The OAM of a magnetic and geometric singularity in the $n$th Landau level is given by the moment formula 

\begin{align}
L_{tip}^{(n)} = &\int \left( \frac{r^{2}}{2} - 2n - 1\right) \left( \sum_{m = 0}^{\infty} |\Psi_{s m + a}^{(n)}|^{2} - \frac{1}{2\pi }\right) dV \label{OAM_hLL1}\\
&- \sum_{n' = 0}^{n-1} \int \left( \frac{r^{2}}{2} - 2n' - 1\right) |\Psi_{-\lambda(n - n') + a}^{(n')}|^{2} dV	\label{OAM_hLL2}\\
& = \frac{c_{n} - 12 \nu \bar{s}_{n}^{2}}{24} \left( s - s^{-1}\right) +  \frac{1}{2}a\left( 2\bar{s}_{n} - \frac{a}{s}\right)\label{OAM_hLL3},
\end{align}
where $c_{n} = 1$ and $\bar{s}_{n} = n + 1/2$. 

Let the energy $E = \hbar \omega_{B} \left( M + \frac{1}{2}\right)$. For fixed $M$, there will be infinitely many states at $n = M$ and $k > 0$. The density of states computed from just these eigenstates will approach a constant $1/2\pi $ away from the cone tip. This is what appears in the first integral in Eqn. (\ref{OAM_hLL1}), and accounts for the subtraction by the asymptotic density.

For $s = 1$ and $a = 0$,  in addition to these positive $k$ states, there will be $M$ additional states with negative $k < 0$ which are degenerate with the $n = M$ states. These appear in the second integral in Eqn. (\ref{OAM_hLL2}). They will be labeled by different $n$ indices, although they belong to the same LL.

As a consequence of the kinetics on a conical singularity, the moment formula which computes the topological contribution to the OAM of a defect will in principle involve non-degenerate states. For non-integer $s$, these are midgap states which exist between degenerate LLs. However, for integer $s$, they become degenerate with a higher $LL$ than they started with. So for instance the state with $(n, m) = (0, -1)$ will have $M =  n + s $ according to (\ref{energy}). 

Nevertheless, this mixing is easily accounted for, and we can ultimately write a formula for the shifted second moment of the density of strictly degenerate states at energy $E = \hbar \omega_{B} (n + \frac{1}{2})$. It will read

\begin{align}
\Delta M_{2}^{(n)} = &\int \left( \frac{r^{2}}{2} - 2n - 1\right) \left( \rho_{n} - \frac{1}{2\pi }\right) dV \label{moment} \\
 =& L_{tip}^{(n)}  + (s n - a) n - \frac{1}{2}s n (n-1) \\
& + \sum_{n' = 0}^{n-1} (|k| + 2 n' - 2n) \delta_{|k| + n', n}
\end{align}
where the the OAM of the defect is

\begin{align}
L_{tip}^{(n)} = \frac{c_{n} - 12 \nu \bar{s}_{n}^{2}}{24} \left( s - s^{-1}\right) +  \frac{1}{2}a\left( 2\bar{s}_{n} - \frac{a}{s}\right)	
\end{align}

The last term accounts for the accidental degeneracy due to an itinerant level. 

\bigskip

{\bf Fixing Experimental Parameters} Here we set $s = 3$, and consider the appropriate form of the sum rule for $a \in \{ 0, 1, 2\}$. 

For $a = 0$ and $s = 1$ the quantum numbers $(n, m)$ which label a state at a given energy $E \sim  M$ are

\begin{align}
M = 0, &\quad (0, 0), (0, 1), ....\\
M = 1, &\quad (0, -1); (1,0), (1,1), ...\\
M = 2, &\quad (0,-2), (1, -1); (2, 0), (2, 1), ...	\\
M = 3, &\quad (0,-3), (1,-2), (2, -1); (3, 0), (3, 1), ...	
\end{align}

On a cone with $s = 3$, if we keep these original labels, the states which remain degenerate with a given Landau level are color coded below for ${\red a= 0}$, ${\blue a = 1}$, and $a = 2$:

\begin{align}
M = 0, &\quad {\red (0, 0)}, {\blue (0, 1)}, (0,2), {\red (0,3)}, ...\\
M = 1, &\quad (0, -1); {\red (1,0)}, {\blue (1,1)}, (1,2), {\red (1,3)}, ...\\
M = 2, &\quad {\blue (0,-1)}, (1, -1); {\red (2, 0)}, {\blue (2, 1)},(2,2),   ...	\\
M = 3, &\quad {\red (0,-1 )}, {\blue (1,-1)}, (2, -1); {\red (3, 0)}, {\blue (3, 1)}, (3,2),  ...	
\end{align}

For example, the lowest Landau level will consist of states with $n = 0$ and $m = a + 3 p$, for $p = 0, 1, ..., $. Furthermore, we can see from this graphic that for $a = 0$, the state with $(0,-1)$ becomes degenerate with the fourth LL. Evaluating the moment formula Eqn. (\ref{moment}) gives us the first few moments relevant for the experiment

\begin{align}
\Delta M_{2}^{(0)} = &L_{tip}^{(0)}\\
\Delta M_{2}^{(1)}  = &L_{tip}^{(1)}  + (s  - a) + \delta_{q,2} \left(s - a  - 2\right) \\
\Delta M_{2}^{(2)} = &  L_{tip}^{(2)} + 3 s - 2a + \delta_{a,1} \left( s  - a  - 4\right)\\
& + \delta_{a,2} \left( s  - a -2\right) .
\end{align}

\begin{figure}[t]
\includegraphics[width=\columnwidth]{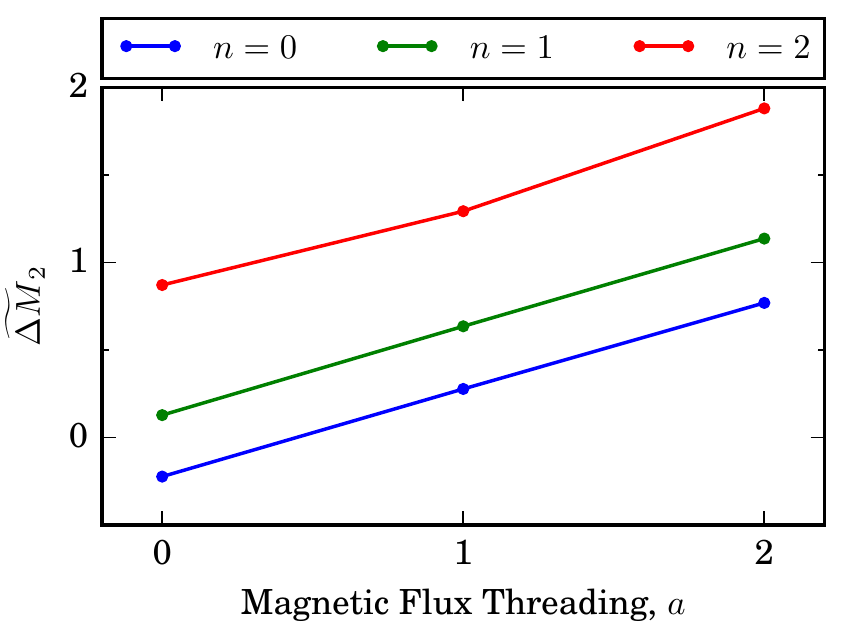}
\caption{\textbf{Extracting $\bar{s}$ from $\Delta M_2$.} We plot $\Delta M_2 (a)$ in the lowest (blue), first excited (green), and second excited (red) Landau levels. The average slope of each line provides a measurement of the mean orbital spin.\label{Figure:dM2slopes}}
\end{figure}

\subsection{Extracting $\bar{s}$ From Density Distribution} 

From the form of the previous equations, it is clear that the only piece of $\Delta M_2$ that depends linearly on $a$ takes the form $(\bar{s}_n-n)a$. To take advantage of this in order to make an independent measurement of the mean orbital spin, we define
\begin{equation}
\widetilde{\Delta M}_2^{(n)} = \Delta M_2^{(n)} - 
			\begin{cases}
			a^2/2s & n=0 \nonumber \\
            a^2/2s +\delta_{a,2} & n=1 \nonumber \\
            a^2/2s +2\delta_{a,1}+\delta_{a,2} & n=2
			\end{cases}
\end{equation}
from which it follows that $\frac{\partial}{\partial a}\left[\widetilde{\Delta M}_2^{(n)}\right] = \bar{s}_n-n$. Since an integer quantum Hall state is expected to have $\bar{s}_n = n+\frac{1}{2}$, we expect the slope $\frac{\partial}{\partial a}\left[\widetilde{\Delta M}_2^{(n)}\right]$ to be independent of Landau level excitation number and equal to $\frac{1}{2}$. In Fig. \ref{Figure:dM2slopes}, we plot $\widetilde{\Delta M}_2^{(n)}(a)$ and observe a equal slopes within each Landau level. We thus measure $\bar{s}_n-n = \{0.496(4),0.504(3),0.505(68)\}$ for $n=\{0,1,2\}$.

\section{Connection of $\delta N$ to $\bar{s}$}
\label{SI:dNtoSbar}
Finally, we return to the charge on a cone tip in a higher Landau level. Using similar arguments as above to account for mid-gap states jumping between Landau levels and resulting in accidental degeneracies, we have that in terms of the density of degenerate states $\rho_{n}$ at energy $E = \hbar \omega_{B} (n + \frac{1}{2})$,

	\begin{align}
\delta N = \int \left( \rho_{n} - \frac{1}{2\pi}\right) dV =& Q_{tip}^{(n)} - n	 + \sum_{n' = 0}^{n-1} \delta_{|k| + n', n}
\end{align}

where
\begin{align}
Q_{tip}^{(n)} & = - \frac{a}{s} + \bar{s}_{n} \left( 1- s^{-1}\right)
\end{align}
is what follows from the topological action according to Eqn.(\ref{Qtip-action}). For the fluxes considered in this experiment, we have

\begin{align}
\delta N &= Q_{tip}^{(0)}\\
\delta N & = Q_{tip}^{(1)}-1+ \delta_{a,2}\\
\delta N & =  Q_{tip}^{(2)}-2	 + \delta_{a, 1} + \delta_{q,2}
\end{align}

The Kronecker delta functions account for the accidental degeneracies that may arise upon tuning the flux.

\end{document}